\documentclass{aa}  

\usepackage{graphicx}
\usepackage{txfonts}
\begin{document} 
   \title{New asteroid clusters and evidence of collisional fragmentation in the $\mbox{L}_{5}$ Trojan cloud of Mars}
 \titlerunning{Martian Trojan clusters}
   \author{A. A. Christou
          \inst{1,2}
          \and
          N. Georgakarakos\inst{3,4}
          \and
          M. \'{C}uk\inst{5}
          \and
          A. Dell'Oro\inst{6}
          \and
          A. Marshall-Lee\inst{1,7}
          \and
          A. Humpage\inst{1,7}
          }
   \institute{Armagh Observatory and Planetarium, College Hill, Armagh BT61 9DG, United Kingdom\\             \email{apostolos.christou@armagh.ac.uk}
            \and  Department of Astronomy, University of Florida at Gainsville, Gainsville, FL 326112055, USA
         \and
             Division of Science, New York University Abu Dhabi, Abu Dhabi, UAE  
          \and  Center for Astro, Particle and Planetary Physics ($\mbox{CAP}^3$), New York University Abu Dhabi, UAE 
          \and  SETI Institute, Mountain View, CA 94043, USA   
          \and  INAF, Osservatorio Astrofisico di Arcetri, I-50125 Firenze, Italy
          \and Department of Physics and Astronomy, Queen's University Belfast, University Road, Belfast, BT7 1NN, Northern Ireland, UK
          \\
             }
   \date{Received January 17, 2025; accepted MMMM DD, 2025}
 
  \abstract
   {Trojan asteroids of Mars date from an early phase of solar system evolution. Based on a sample of $<$10 asteroids, MT distribution has been previously shown to be both asymmetric and inhomogeneous with population evolution dominated by thermal radiation forces acting over timescales of Gyr. Remarkably, a single asteroid family associated with (5261) Eureka ($H$$\sim$16) at $\mbox{L}_{5}$ contains most MTs and all members of the stable population fainter than $H$=18.}
   {Using the currently available sample of MTs and their orbits, we take a fresh look at this population to re-evaluate these earlier conclusions and to search for additional features diagnostic of the evolutionary history of MTs and the Eureka family in particular.}
   {We perform harmonic analysis on numerical time series of the osculating elements to compile a new proper element catalogue comprising 16 $\mbox{L}_{5}$ and 1 $\mbox{L}_{4}$ Mars Trojan asteroids. We then combine sample variance analysis with statistical hypothesis testing to identify clusters in the distribution of orbits and assess their significance.}
   {We identify two small clusterings significant at 95\% confidence of three $H$=20-21 asteroids each and investigate their likely origin. One of the clusters is probably the result of rotational breakup of a Eureka family asteroid $\sim$$10^{8}$ yr ago. The significantly higher tadpole libration width of asteroids in the other cluster is more consistent with an origin as impact ejecta from Eureka itself and on a timescale comparable to the $\sim$1 Gyr age of its family. We further confirm the previously reported correlations in Eureka family orbital distribution attributed to the long-term action of radiation-driven forces and torques on the asteroids.}
   {}
   \maketitle
%
\section{\label{sec:intro} Introduction}
All major planets in the solar system except Mercury sport asteroidal companions in the so-called tadpole or Trojan configuration, where a particle is confined to the vicinity of the Lagrangian triangular equilibrium locations that lead or trail the planet by $60^{\circ}$ \citep{MurrayDermott1999}. Numerical simulations demonstrate the transient nature of many of these Trojans and, by implication, their likely origin as dynamically unstable planet-crossing small bodies \citep{Christou2000a,MoraisMorbidelli2002,Connors.et.al2004}.

Trojans of Jupiter, Neptune and Mars have Gyr dynamical lifetimes and likely date from events early in Solar System history \citep{Levison.et.al1997,Scholl.et.al2005,LykawkaHorner2010} with Mars featuring the only set of dynamically long-lived Trojans among terrestrial planets. Unlike Main Belt asteroids, the Mars Trojans (MTs hereafter) experience a relatively benign collisional environment, while their proximity to the Sun and long dynamical lifetime make them a useful proving ground for models of asteroid orbital-rotational evolution by radiation forces \citep{Rivkin.et.al2003,Cuk.et.al2015,Christou.et.al2017}.

Their small sizes notwithstanding \cite[$\lesssim$2 km;][]{Trilling.et.al2007}, MTs exhibit some unique features: a marked asymmetry favouring $\mbox{L}_{5}$ vs $\mbox{L}_{4}$ residents \citep{deLaFuenteMarcoses2013,Christou2013}; a strong orbital concentration near (5261) Eureka, suggesting that most $\mbox{L}_{5}$ Trojans are products of YORP-induced rotational spin-up and breakup \cite[``YORPlets'';][]{Christou.et.al2017}; and the olivine-dominated composition of the family \citep{Borisov.et.al2017,Polishook.et.al2017}, one of only two known to exist \citep{Galinier.et.al2024}. 

The Minor Planet Center website\footnote{https://minorplanetcenter.net/iau/lists/MarsTrojans.html} lists 16 Mars Trojan asteroids, of which all but two have been confirmed as stable residents by numerical integration of the orbit \citep{Scholl.et.al2005,Christou2013,deLaFuenteMarcoses2013,Cuk.et.al2015,Christou.et.al2020,delafuentemarcos2021}. Of the two exceptions, 2023 $\mbox{FW}_{14}$ is the second $\mbox{L}_{4}$ Trojan of Mars to be discovered \citep{delafuentemarcos2024}. Its short, $\sim$$10^{7}$ yr dynamical lifetime points to a likely origin in the Mars-crosser population. The other asteroid, 2018 $\mbox{FM}_{29}$, was discovered in 2018 and observed again in 2020, 2022 and 2024. 
The orbital semimajor axis uncertainty of this asteroid is now several orders of magnitude smaller than the radial width of the MT region and we have confirmed by numerical integration its status as a stable resident at $\mbox{L}_{5}$. Similarly, the recovery of single-opposition asteroids 2015 $\mbox{TL}_{144}$ and 2016 $\mbox{AA}_{165}$ in 2022 and 2024 respectively place them firmly within the Trojan clouds of Mars. Inclusion of the new Trojans raises the tally of MTs to 18 and of $\mbox{L}_{5}$ residents in particular to 16. 
\begin{figure}
   \centering
   \includegraphics[width=\columnwidth]{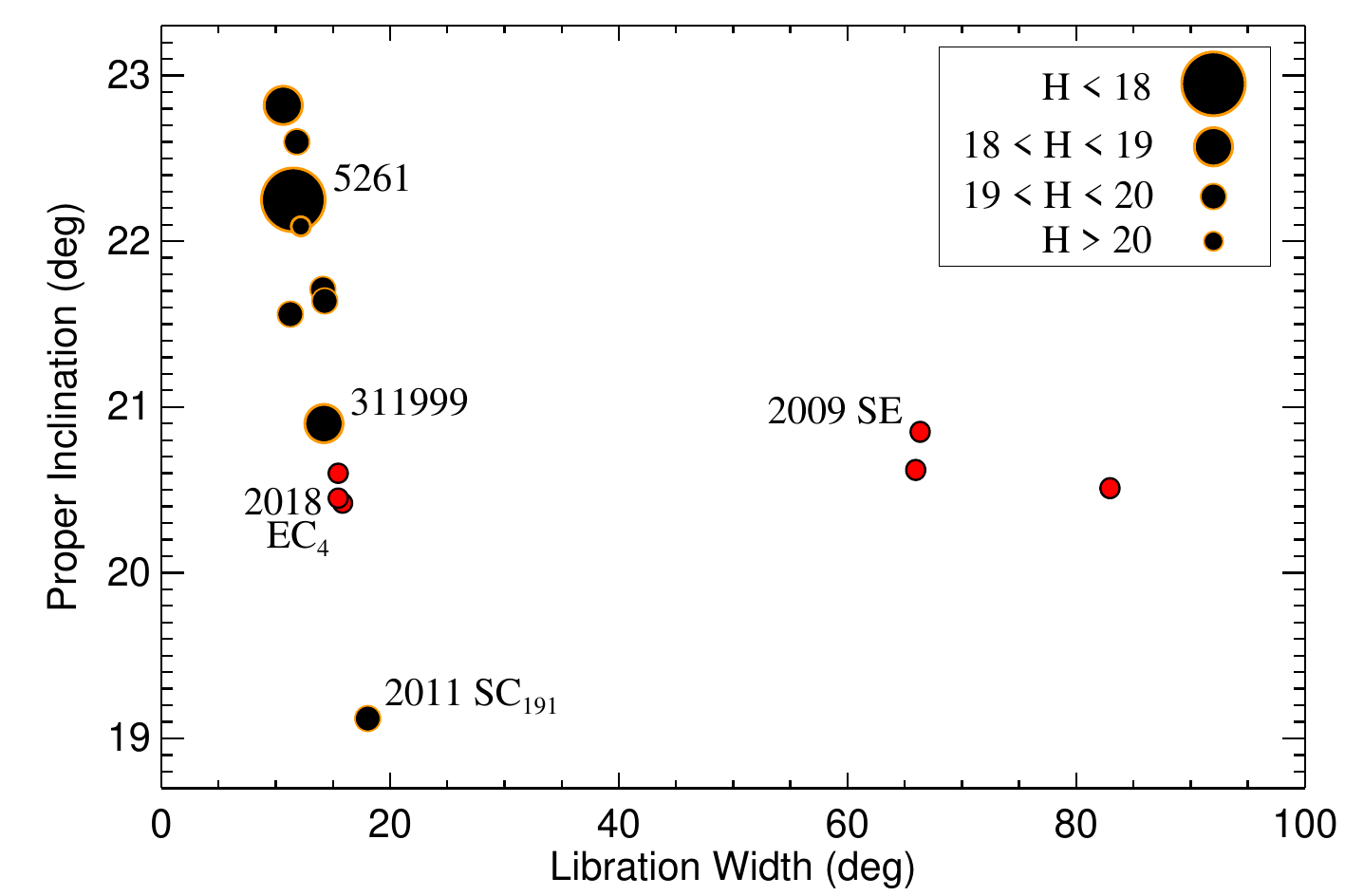}
   \includegraphics[width=\columnwidth]{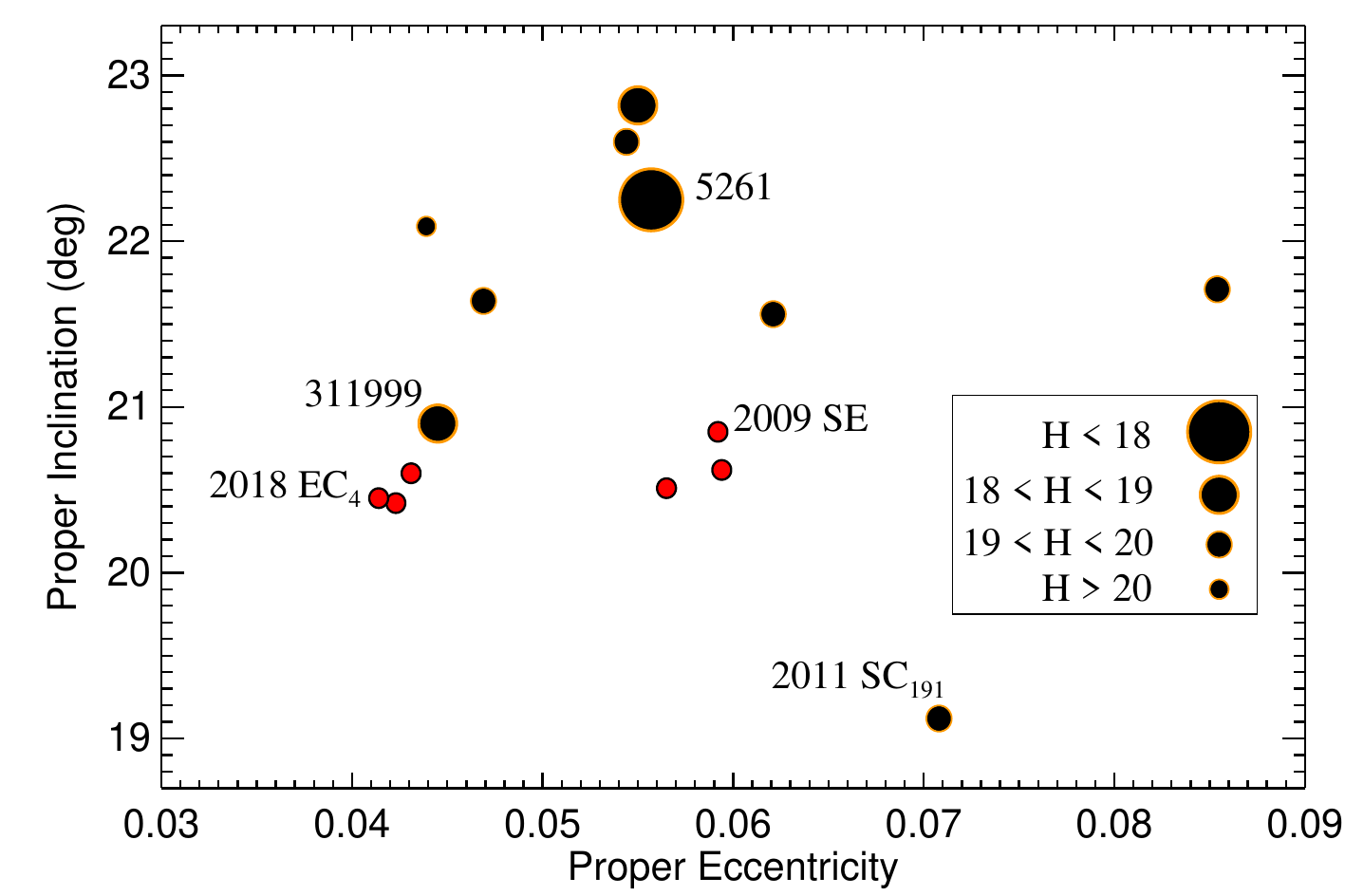}
      \caption{Top: Distribution of $\mbox{L}_{5}$ Mars Trojans in libration width ($L$) vs inclination ($I_{P}$) space (top) and eccentricity $e_{P}$ vs inclination $I_{P}$ (bottom). Red points correspond to new asteroid clusters identified in this work. 
      } 
         \label{fig:lr_vs_e_vs_i}
   \end{figure}  
   
Using the methods described in the following Section of the paper, we have calculated proper elements - libration width $L$, eccentricity $e_{P}$ and inclination $I_{P}$ - (Table~\ref{table:propelems}) from the latest osculating elements of the objects available through the {\tt AstDys}\footnote{{\tt https://newton.spacedys.com/astdys2/}. Data retrieved 20 September 2024} online service. Their distribution, shown in Fig.~\ref{fig:lr_vs_e_vs_i}, shows several interesting features. A strong correlation between resonant libration width and inclination among Eureka family asteroids was noted previously by \citet{Cuk.et.al2015} and shown there to be a consequence of the solar-thermal Yarkovsky effect \citep{Bottke.et.al2006}. \citeauthor{Cuk.et.al2015} further noted that the asymmetric orbital distribution of family Trojans with respect to Eureka is compatible with a non-gravitational acceleration acting {\it against} the direction of motion, as might arise if bi-directional diurnal Yarkovsky is suppressed relative to the strictly dissipative seasonal variant through the coupling of spin and shape evolution \citep{Rubincam2000,VokrouhlickyCapek2002,CapekVokrouhlicky2004,Statler2009,CottoFigueroa.et.al2015}. Secondly, the new MTs have eccentricity and inclination similar to the Eureka family but much higher libration width. Their orbits are similar to that of 2009 SE, an $\mbox{L}_{5}$ Trojan that might be unrelated to Eureka \citep{delafuentemarcos2021}.

The availability of a larger MT sample than was available in earlier works
motivated us to take a fresh look at their orbital distribution using tools from dynamical systems and from statistical analysis. The present study has uncovered evidence for a more complex evolutionary history of this population than recognised previously, including secondary rotational break-up of YORPlet asteroids and a role for collisions as a competing process for MT population evolution alongside thermal-rotational effects. The next Section exposes the methods used to compile a new catalogue of MT proper elements, extract interesting features from it and evaluate their statistical significance. Our results are described in Section~\ref{sec:results} while in Section~\ref{sec:origin} we examine origin scenarios for the new clusters.  Section~\ref{sec:conclusions} presents our conclusions. 
\begin{figure*}
   \centering
   \includegraphics[width=2.1\columnwidth]{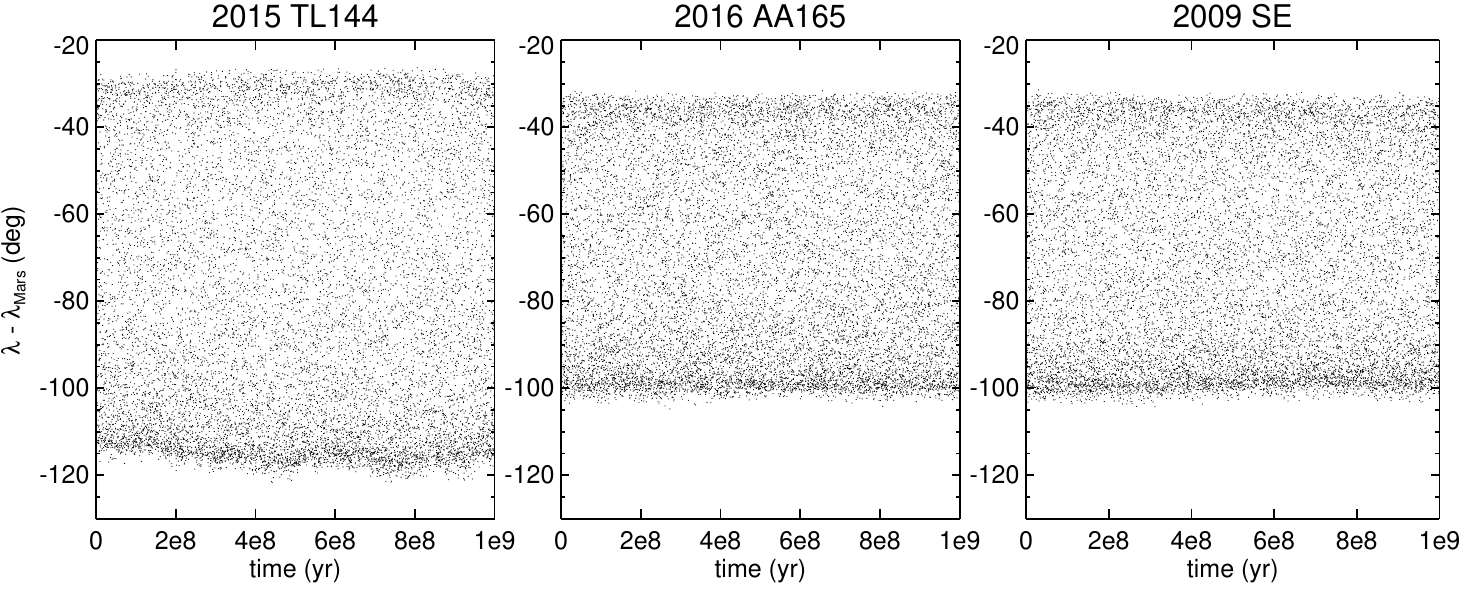}
      \caption{Critical angle evolution over $10^{9}$ yr for dynamical clones of selected Mars Trojan asteroids.} 
         \label{fig:onegyr}
   \end{figure*}
\section{\label{sec:methods}Methods}
\subsection{\label{sec:stability}Confirming the stability of Mars Trojans 2009 SE, 2015 $\mbox{TL}_{144}$, 2016 $\mbox{AA}_{165}$ and 2018 $\mbox{FM}_{29}$}
To determine the long-term stability of candidate Trojans, 50 dynamical clones of each asteroid were generated from the state covariance at epoch MJD 60600 available through {\tt AstDys} using the method of \citet{Duddy.et.al2012}. The clones plus the nominal orbit were then numerically integrated for $10^{9}$ yr with the MERCURY package \citep{Chambers1999} under the gravity of the eight major planets and an integration time step of 4d. For all clones of each asteroid, the critical angle $l_{r}$=$\lambda - \lambda_{\rm Mars}$ remained in libration around $-60^{\circ}$ until the end of the integration, indicating dynamical lifetimes comparable to the solar system age. Examples of individual clone orbital evolution are shown in Fig.~\ref{fig:onegyr} where we have also included 2009 SE. In the case of 2015 $\mbox{TL}_{144}$ we observe libration width variations of $5^{\circ}-10^{\circ}$ over timescales of $10^{8}$ yr. Such variations are seen in most clones of the asteroid and we attribute them to the proximity of the orbit to the chaotic region surrounding the tadpole-horseshoe separatrix at a libration width of $\sim$$156^{\circ}$ \citep{Morais1999}. Otherwise, the clone integrations imply similar dynamical stability to the other Trojans. 
\begin{figure}
   \centering
   \includegraphics[width=\columnwidth]{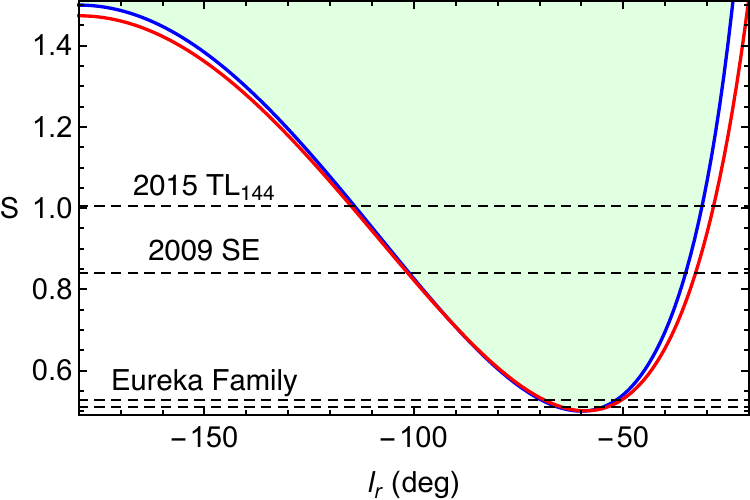}
      \caption{Profiles of the effective potential (Eq.~\ref{eq:sdef}) for tadpole-type libration (green) evaluated with $I=0^{\circ}$ (blue curve) and $I=20^{\circ}$ (red curve) to highlight the effect of orbital inclination on libration width.} 
         \label{fig:potential}
   \end{figure}   
\subsection{\label{sec:theory}Modelling tadpole libration of Mars Trojans}
In the context of the circular restricted Sun-planet-asteroid problem, the evolution of the guiding centre $a_{r}(l_{r})$ in the vicinity of the planetary orbit ($|a_{r}|= |a - a_{P}| \ll$1) is described by the integral \citep{Namouni.et.al1999}
\begin{eqnarray}
\label{eq:jacobi_int}
 {a}^{2}_{r}& =& C - \frac{8 \mu}{3} S(l_{r},e, I,\omega)\\
 \label{eq:sdef}
 S\left(l_{r}\mbox{, }e\mbox{, }I\mbox{, }\omega \right)& =&\left. \frac{1}{2 \pi}\int^{\pi}_{-\pi}{\left({|\mathbf{r} - \mathbf{r}_{P}|}^{-1} - \mathbf{r}\cdot\mathbf{r}_{P} \right)}\mbox{ }\mathrm{d}\mathrm{\lambda}  \right|_{a_{r}=0}
\end{eqnarray}
where $C$ is a constant, $\mathbf{r}$ and $\mathbf{r}_{P}$ are the heliocentric position vectors of the asteroid and planet respectively, $l_{r}=\lambda-\lambda_{P}$ and distances are in units of $a_{P}$. As $S$ has a minimum $S_{\rm min}$$\simeq$$1/2$ at $l_{r}$$\simeq$$\pm$$\pi/3$, Eq.~\ref{eq:jacobi_int} can be written as
\begin{equation}
\label{eq:jacobi_new}
 {a}^{2}_{r} = {a}^{2}_{0} - (8 \mu/3) \mbox{ }\Delta S
\end{equation}
where $\Delta S = S - S_{\rm min}$ and ${a}_{0}^{2} = C - \left(8 \mu/3\right)S_{\rm min}$. The radial width $a_{0}$ for tadpole libration is obtained by evaluating Eq.~\ref{eq:jacobi_new} at the turning points for $0$$<$$\Delta S $$< 1$. For $e\sim 0\mbox{, }I\sim0^{\circ}$ and $\Delta S$=1 we obtain the classical result \citep{Yoder.et.al1983,MurrayDermott1999,Morais1999,Morais2001} that
\begin{equation}
{a}_{0,\rm max} \simeq \sqrt{\left(8/3\right)\mu} \mbox{ } a_{P}\mbox{.}
\label{eq:libration_width}
\end{equation}
For the case of Mars, we use $\mu$$\sim$$ 3.2\times 10^{-7}$, $a_{\rm P}$$\sim$$ 1.524$ au and find
\begin{equation}
\label{eq:mars_width}
a_{0,\rm max} \simeq 0.00141  \mbox{ au.}
\end{equation}
At large libration amplitude, the motion of the guiding centre around the equilibrium point is strongly asymmetric \citep{Erdi1988,Milani1994}. For this reason, we opt to parameterise the resonant state of the asteroid orbit by the full width of $l_{r}$ libration rather than the libration amplitude $D$ \citep{Milani1993,Cuk.et.al2015} and refer to this width as $L$ so that, for small libration amplitude ($S$$\gtrsim$$S_{\rm min}$), $L$$\simeq$$2D$. We also choose $\bar{d}={a}_{0}/{a}_{0,\rm max}=\Delta S$ as the action-like quantity, obtained by numerical evaluation of Eq.~\ref{eq:jacobi_new} at $a_{r}$$=$$0$ with $\omega=0^{\circ}$, $e=0$ and $I=0^{\circ}$. Our approach in determining the resonant elements is similar to that of \citet{Vokrouhlicky.et.al2024} for the Jupiter Trojans. Values of the proper quantities for the Mars Trojans are reported in Table~\ref{table:propelems}. It is worth pointing out that the moderate inclination of MTs modifies the effective potential with respect to the planar case (Fig.~\ref{fig:potential}). Its principal effect is a planetward displacement of the turning point of the tadpole by up to $\sim$$4^{\circ}$ and an increase in $L$ by a similar amount. This effect manifests as a fixed offset in the libration widths of the clusters investigated in Section~\ref{sec:clusters}. Here we have chosen not to include this correction in our computation of $L$ to maintain consistency with existing proper element catalogs \cite[eg][]{Vokrouhlicky.et.al2024}. 
\subsection{\label{sec:proper}Proper element estimation}
We calculate the proper eccentricity and inclination by carrying out a $\sim$7.3 $\times$ $10^{5}$ yr integration of the nominal orbit as in Section~\ref{sec:stability} with a 4d time step and sampled every 128d. The complex quantities $ z = e \exp {\rm i} \varpi $ and $\zeta = I \exp {\rm i} \Omega $ formed from the output are then fitted to Fourier series of the form
\begin{equation}
\label{eq:secular}
z(t)  = A_{0} e^{{\rm i} g_{0} 
t} + \sum^{n}_{i=1}A_{i} e^{\rm i \nu_{i}  t}\mbox{, }\zeta(t)  = B_{0} e^{\rm i s_{0} t}+\sum^{n}_{i=1} B_{i} e^{\rm i \mu_{i} t}
\end{equation}
 using the Frequency-Modified Fourier Transform algorithm \citep[FMFT;][] {SidlichovskyNesvorny1997}. The coefficients $A_{i}$ and $B_{i}$ in Eq.~\ref{eq:secular} are complex quantities. The frequencies $\nu_{i}$, $\mu_{i}$ with $i \neq 0$ are linear combinations of the fundamental solar system secular eigenmodes \citep{Laskar.et.al2004} while $\nu_{0}$ and $\mu_{0}$ are the free precession eigenmodes of the asteroid and ${\rm i} = \sqrt{-1}$. Proper elements are recovered from the fit as $e_{P}$=$|A_{0}|$ and $I_{P}$=$|B_{0}|$. 
 
\begin{figure}
   \centering
   \includegraphics[width=0.95\columnwidth]{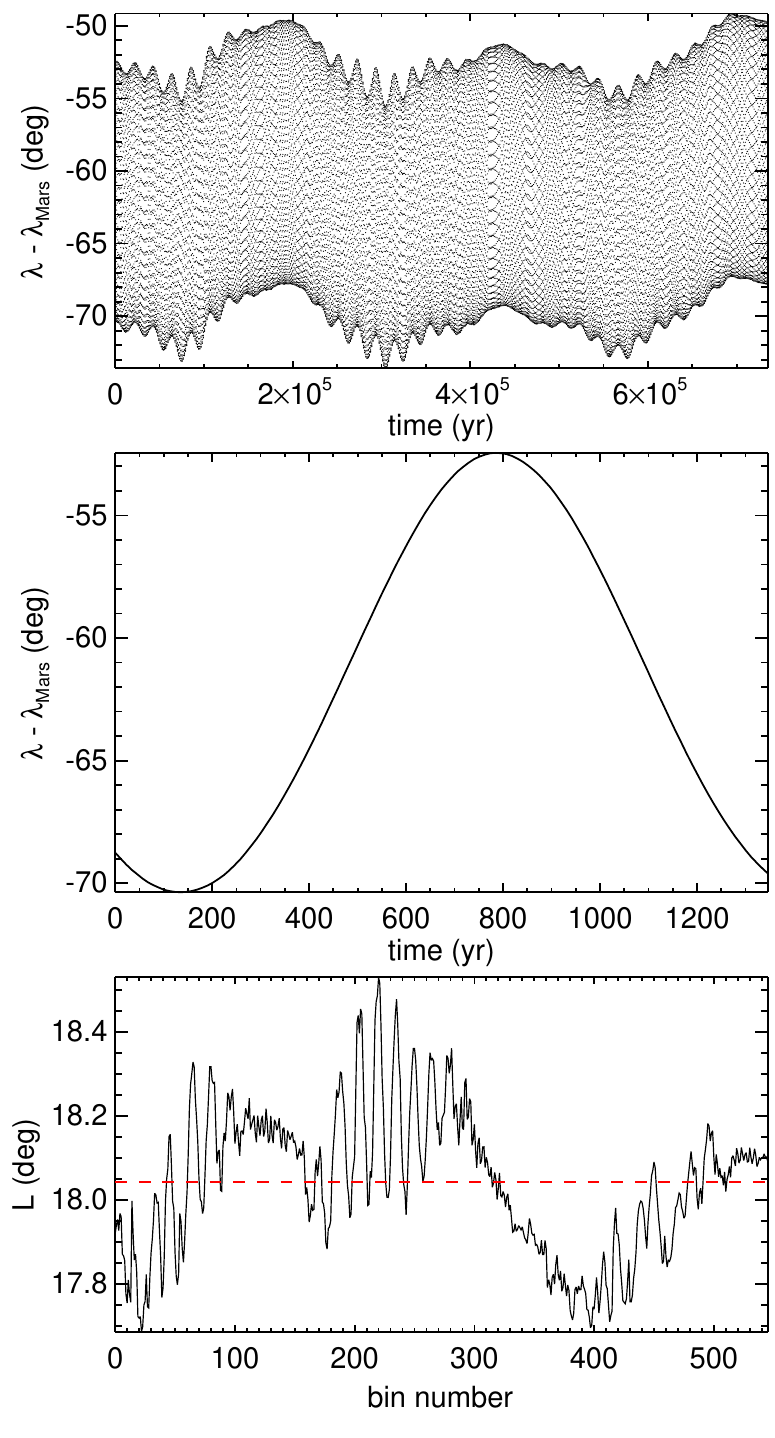}
      \caption{Example of libration width estimation for Mars Trojan 2011 $\mbox{SC}_{191}$ from the numerical simulations. Top: Evolution of the critical angle $l_{r}$. Middle: Variation of $l_{r}$ over one boxcar width of 120 samples or $\sim$1,350 yr. Bottom: Variation of the $L$ estimate over all boxcar windows with the average value (red dashed line) reported in Table~\ref{table:propelems}.} 
         \label{fig:lwunc}
   \end{figure}

 For the libration width $L$ we adopt the average of the difference between the maximum and minimum value of $l_{r}$ within a 120-sample window applied to the numerical time series. An example of applying this procedure is shown in Fig.~\ref{fig:lwunc}. Finally, we obtain $\bar{d}$ from $L$ by evaluating Eq.~\ref{eq:jacobi_new} for a circular, planar orbit as described in the previous Section.
\begin{table}
\caption{Proper elements of Gyr-stable Mars Trojan asteroids.}             
\label{table:propelems}      
\centering          
\begin{tabular}{@{}l@{ }ccccc}     
\hline\hline     \noalign{\smallskip}   
 & $L$ & $\bar{d}$ &  & $I_{P}$ &  \\ 
Designation & (deg) & (au) & $e_{P}$ &  (deg)& $H$ \\ 
\hline \noalign{\smallskip}               
(5261)  & 11.53 (13) & 0.0114 (02) & 0.0557 & 22.25  &  16.17 \\ 
(101429)   & 45.38 (70) & 0.1692 (51) & 0.0749 & 32.17  & 17.27 \\
(121514)   &  78.13 (60) & 0.4567 (59) & 0.0378 & 18.05  & 17.18 \\
(311999)  &  14.21 (13)  & 0.0172 (03)  & 0.0445 & 20.90  & 18.12  \\
(385250)  & 10.65 (13) &  0.0097 (02) & 0.0550  &  22.82  &  18.86 \\ 
2009 $\mbox{SE}$  &  66.35 (72) & 0.3421 (80) & 0.0592 & 20.85 & 20.04  \\
2011 $\mbox{SC}_{191}$ & 18.04 (17) &  0.0277 (05) & 0.0708  &  19.12  & 19.54 \\
2011 $\mbox{SL}_{25}$ & 14.12 (18) & 0.0170 (04) &  0.0854  &  21.71 &  19.49  \\
2011 $\mbox{SP}_{189}$ & 15.84 (38) & 0.0214 (10) &  0.0423 & 20.42  & 20.96  \\
2011 $\mbox{UB}_{256}$ & 11.84 (13) & 0.0120 (02) & 0.0544  &  22.60  &  19.46  \\
2011 $\mbox{UN}_{63}$ & 14.28 (13) & 0.0174 (03)  &  0.0469  &  21.64   &   19.81 \\
{\bf 2015} $\mathbf{\mbox{\bf TL}_{144}}$ & {\bf 82.95 (76)} & {\bf 0.5047 (76)} & {\bf 0.0565} &  {\bf 20.51} & {\bf 20.99} \\
{\bf 2016} $\mathbf{\mbox{\bf AA}_{165}}$ & {\bf 65.96 (63)} & {\bf 0.3397 (58)} & {\bf 0.0594} & {\bf 20.62} & {\bf 20.37} \\
2016 $\mbox{CP}_{31}$  & 11.27 (13) & 0.0108 (02) & 0.0621 & 21.56 &  19.59  \\
2018 $\mbox{EC}_{4}$  &  15.46 (15) & 0.0204 (04) & 0.0414 & 20.45  & 20.16  \\ 
2018 $\mbox{FC}_{4}$  & 12.31 (12) & 0.0129 (03) &   0.0439 & 22.09  & 21.30  \\ 
{\bf 2018} $\mathbf{\mbox{FM}_{29}}$ & {\bf 15.46 (15)} & {\bf 0.0204 (04)} &  {\bf 0.0431} & {\bf 20.60}  & {\bf 21.15} \\
 \noalign{\smallskip}
\hline                  
\end{tabular}
\tablefoot{Uncertainties reported at 95\% confidence. The second column is the libration full width accompanied by the formal uncertainty in units of $10^{-2}$ degrees. The third
column is the equivalent width in semimajor axis normalised to $a_{\rm 0,max}$ and uncertainty in units of $10^{-4}$. The fourth and fifth columns are the proper eccentricity and inclination with formal uncertainty of $1.5\times10^{-3}$ and $0.1^{\circ}$ respectively. Entries shown in bold correspond to objects confirmed as stable Mars Trojans in this work. Error modelling details are provided in Appendix Section \ref{app:propunc}.}
\end{table} 
\begin{figure}
   \centering
   \includegraphics[width=\columnwidth]{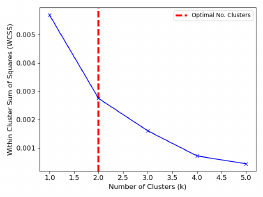}
      \caption{The WCSS as a function of the cluster number $k$ in {\tt kmeans}. The elbow (vertical dashed line) identifies the optimum value $k=2$.} 
         \label{fig:elbow_method}
   \end{figure}
\subsection {\label{sec:stats}Statistical tests of cluster significance}
To establish that any particular asteroid grouping represents a real association, we use Monte-Carlo simulations to form the distribution of an appropriate test statistic and compare it with its value for the candidate cluster to estimate the probability - or $p$-value - that the claimed cluster is a chance arrangement between unrelated objects. Two different statistical functions are used, one based on the Minimum-volume Enclosing Sphere (MES) and the other on the Cluster Index (CI).

Given a set of $m$ points in a multi-dimensional Euclidean space, the MES of that set is defined by the property that 
\begin{equation}
V(P_{\rm MES})<V(P), \forall P:\{{\mathbf{x}_{i}}\}_{i=1\cdots m}\in P 
\end{equation} 
where $P(\mathbf{x}_{0},r_{0})$ is the sphere with centre $\mathbf{x}_{0}$ and radius $r_{0}$ 
while $V(P)$ is the sphere volume, used here as a measure of compactness for a given $m$-subset of $M$ proper orbits. 

On each MC trial, we generate a synthetic set of M random points from some reference distribution representing the null hypothesis of uniformity. We then calculate the MES volume $V(P_{\rm MES})$ for every possible $m$-subset of points and compare it with the MES volume for the candidate asteroid cluster $V({P_{{\rm MES},c}})$=$V_{c}$, recording  a ``hit'' when the condition $\min\limits_{j}V(P_{j})<\eta \mbox{ }V_{c}$ is satisfied, where $j$ runs through the set of subsets. The quantity 
\begin{equation}
p(M,m)=\frac{\mbox{Number of Hits}}{\mbox{Number of Trials}}
\end{equation} 
then estimates the probability that a cluster as compact as our candidate cluster could have arisen by chance. The $\eta$ factor is introduced as a convenient way to incorporate the different sources of uncertainty in MES volume calculation. In this paper we use $\eta=3.0$ as a compromise between applying an overly strict criterion and the risk of underestimating $V_{c}$ due to the statistical uncertainty in the proper orbits.

The Cluster Index \cite[][see also \citet{Markwardt.et.al2023}]{Liu.et.al2008,Huang.et.al2015} is  defined as
\begin{equation}
\label{eq:ci}
CI=\frac{\sum^{N}_{j=1} \sum_{i \in C_{j}} {||\mathbf{x}^{(j)}_{i}-\mathbf{\bar{x}}^{(j)}||}^{2}}{\sum^{M}_{i=1}{||\mathbf{x}_{i}-\mathbf{\bar{x}}||}^{2}}
\end{equation}
where $\mathbf{\bar{x}}^{(j)}$ is the mean for the $j$-th out of $N$ clusters within a sample of $M$ points and $\mathbf{\bar{x}}$ is the overall sample mean. A CI value $\ll$1 implies a strongly-clustered set. The synthetic set of orbits in each Monte Carlo trial is split into $N$$=$$2$ clusters (of size $M_{1}$ and $M_{2}$ respectively) by the {\tt kmeans} unsupervised clustering search algorithm \citep{MacQueen1967}. The CI values for the randomly-generated sets form the cumulative distribution function to be compared with the CI value for the candidate asteroid cluster so that the $p$-value is finally estimated in a manner analogous to the MES method, ie 
\begin{equation}
p(M_{1},M_{2})=\frac{\mbox{Number of Hits}}{\mbox{Number of Trials}}\mbox{.}
\end{equation} 
{\tt Kmeans} requires the number of clusters as input. Here we use the Elbow method \citep{Thorndike1953} to heuristically determine the optimum cluster number for our sample by evaluating the numerator in Eq.~\ref{eq:ci} as a function of $k$. This is the Within-Cluster Sum of Squares \cite[WCSS;][]{MacQueen1967} with the property that it vanishes as $k$$\rightarrow$$M$. The evaluation at $k$$=$$2$ (Fig.~\ref{fig:elbow_method}) corresponds to an ``elbow'' - the largest change between consecutive values as well as the largest change between consecutive {\it changes} - as the optimum cluster number in this case. A full listing of our statistical tests and their outcomes is presented in Table~\ref{table:sign}.
\section{\label{sec:results}Results}
\subsection{\label{sec:eureka}Eureka family shape and the role of Yarkovsky}
A notable feature of the Eureka family is
a strong anti-correlation between libration width and proper inclination (Fig.~\ref{fig:lr_vs_e_vs_i}, top panel) attributed to Yarkovsky-driven coupled evolution of $L$ and $\sin{I}$ while the evolution of the eccentricity is dominated by random diffusive growth \citep{Cuk.et.al2015,Christou.et.al2020}. Numerical models of orbital evolution via either mode suggest an approximate family age of 1 Gyr when compared to the observed orbital dispersion \citep{Cuk.et.al2015}. Here we revisit these conclusions with the benefit of a larger sample of family asteroids and improved orbit determination due to longer observational arcs. 
\begin{figure}
   \centering
   \includegraphics[width=\columnwidth]{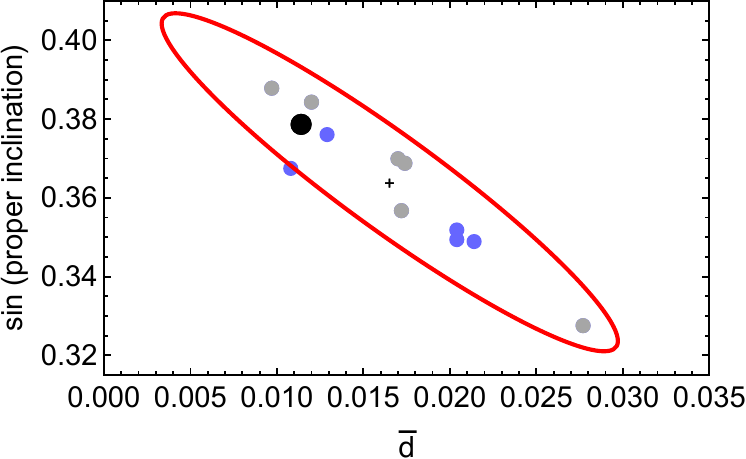}
      \caption{Orbital distribution of Eureka family asteroids. Grey points represent objects considered in \citet{Cuk.et.al2015} while blue points are new Trojans. The black point is (5261) Eureka. The red ellipse demarcates a 95\% confidence region from the variance analysis and the "+" sign marks the mean family orbit.} 
         \label{fig:eureka_shape}
   \end{figure}
The shape and orientation of the family can be computed from the sample covariance matrix
\begin{equation}
\mathbf{\Sigma}(\mathbf{x})=\frac{1}{N-1}
\left(\mathbf{x} - \mathbf{\bar{x}}\right)\left(\mathbf{x} - \mathbf{\bar{x}}\right)^{T}
\end{equation}
where $\mathbf{x}=\{[\bar{d}_{i},\mbox{}e_{i},\mbox{}\sin{I}_{i}]\}_{i=1\cdots M}$ is a $3\times M$ matrix with rows the proper elements for each of the $N$ family members, $\mathbf{\bar{x}}$ is the sample mean and the superscript ``{\it T}'' indicates the transpose.

The eigenvalues and eigenvectors of $\mathbf{\Sigma}$, reported in Table~\ref{table:covelems}, define the shape and orientation of gaussian elliptical iso-contours as shown in Fig.~\ref{fig:eureka_shape}. The coordinate frame defined by the principal axes of the ellipse amounts to a counterclockwise rotation by an angle $\theta=163.7^{\circ}$ from the positive $\bar{d}$ axis. A statistical measure of association between $\bar{d}$ and $\sin{I}$ is the Pearson correlation coefficient 
\begin{equation}
n_{13}=c_{13}/\sqrt{c_{11} c_{33}}
\end{equation}
with values between $+1$ or $-1$. The two extremes correspond to points lying along a line. For the sample of $M=12$ Eureka family members we find 
\begin{equation}
n_{13}=-0.939
\end{equation}
that is, a strong inverse association between $\bar{d}$ and $\sin{I}$. Using only the seven asteroids considered in \citet{Cuk.et.al2015} we obtain the slightly higher negative correlation of $-0.971$.
In contrast, we find no evidence of correlation of either of those elements with $e$ ($|n_{23}|\mbox{, }|n_{12}|<$$0.05$). 

These results reinforce the conclusions of \citet{Cuk.et.al2015} on the controlling influence of non-gravitational forces in the orbital distribution of Mars Trojans. Moreover, all of the five additional Trojans in this study (blue points in Fig.~\ref{fig:eureka_shape}) are located either near Eureka or to higher $\bar{d}$ and lower $\sin{I}$. Therefore, the asymmetric distribution of family Trojans relative to Eureka attributed to the dominance of the seasonal Yarkovsky effect in the 2015 study is also confirmed for our 1.7$\times$ larger sample.
  \begin{figure}
\includegraphics[width=0.9\columnwidth]{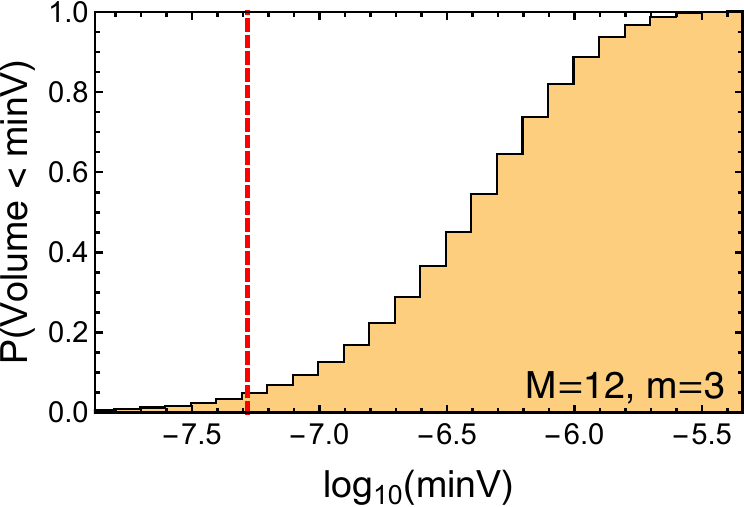} 
   \includegraphics[width=0.9\columnwidth]{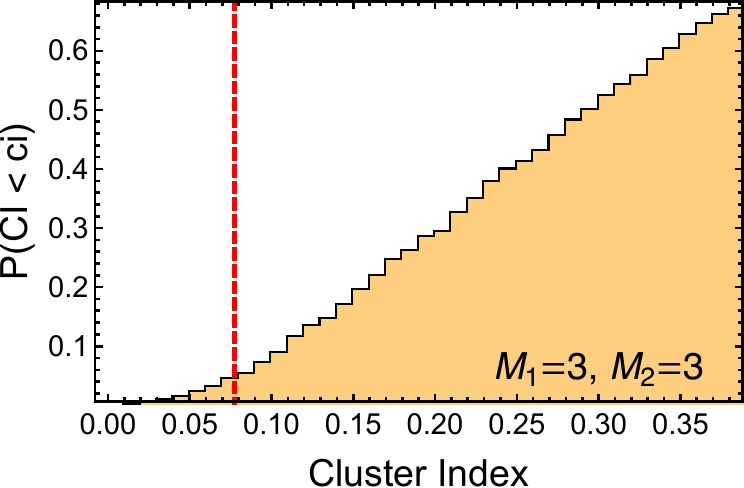}\vspace{1mm}
   \includegraphics[width=0.9\columnwidth]{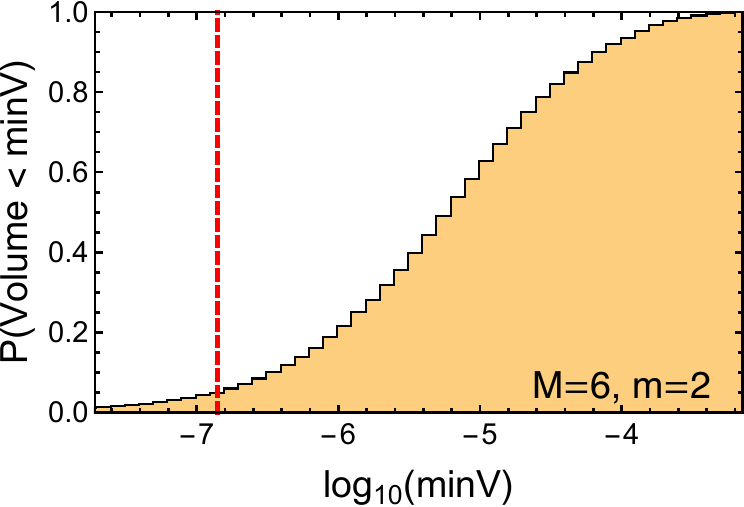}\hspace{2mm}
      \caption{Statistical tests for candidate MT clusters. From top to bottom: MES volume distribution versus the test statistic value (red line) for the 2018 $\mbox{EC}_{4}$ cluster; CI distribution versus the test statistic value for the 2009 SE cluster; and MES volume distribution versus the test statistic value for the 2009 SE $-$ 2016 $\mbox{AA}_{165}$ pair.}
         \label{fig:clustertests}
   \end{figure}
\subsection{\label{sec:clusters}New Mars Trojan clusters}
\begin{table*}
\caption{Significance test results for the Mars Trojan clusters.}
\label{table:sign}      
\centering          
\begin{tabular}{lcc}     
\hline\hline     \noalign{\smallskip}   
Cluster   &  Null Distr. & $p$-value \\\hline \noalign{\smallskip}
 \multicolumn{3}{c}{MES Method} \\ \hline \noalign{\smallskip}
 2018 $\mbox{EC}_{4}$  ($M=12$, $m=3$) & N($\mu$,$\mathbf{\Sigma}$) &  0.042\\ 
\noalign{\smallskip} \cline{1-1} \noalign{\smallskip}
        2018 $\mbox{EC}_{4}$ + (311999) ($M=12$, $m=4$) & N($\mathbf{\mu}$,$\mathbf{\Sigma}$) &  High\\ \noalign{\smallskip} \cline{1-1} \noalign{\smallskip}
  2009 SE ($M=15$, $m=3$) &  N($\mathbf{\mu}$,$\mathbf{\Sigma}$) & High\\ 
\noalign{\smallskip} \cline{1-1} \noalign{\smallskip}
 2009 SE $-$ 2016 $\mbox{AA}_{165}$ ($M=15$, $m=2$) &  $N$($\mathbf{\mu}$,$\mathbf{\Sigma}$)  & 0.086 \\ 
 2009 SE $-$ 2016 $\mbox{AA}_{165}$ ($M=15$, $m=2$) &  U(0,0.6) $\times$ N($\mathbf{\mu}_{23}$,$\mathbf{\Sigma}_{23}$)  & 0.082 \\ 
  2009 SE $-$ 2016 $\mbox{AA}_{165}$ ($M=6$, $m=2$) & $N$($\mathbf{\mu}$,$\mathbf{\Sigma}$)  & 0.044 \\ 
    2009 SE $-$ 2016 $\mbox{AA}_{165}$ ($M=6$, $m=2$) & U(0,0.6) $\times$ N($\mathbf{\mu}_{23}$,$\mathbf{\Sigma}_{23}$)  & 0.020 \\
 2009 SE $-$ 2016 $\mbox{AA}_{165}$ ($M=4$, $m=2$) & $N$($\mathbf{\mu}$,$\mathbf{\Sigma}$)  & 0.005 \\ 
    2009 SE $-$ 2016 $\mbox{AA}_{165}$ ($M=4$, $m=2$) & U(0,0.6) $\times$ N($\mathbf{\mu}_{23}$,$\mathbf{\Sigma}_{23}$)  & 0.005 \\
  \hline     \noalign{\smallskip} 
 \multicolumn{3}{c}{CI Method}\\ \hline     \noalign{\smallskip}  
2009 SE ($M_{1}=15$, $M_{2}=3$) &  N($\mathbf{\mu}$,$\mathbf{\Sigma}$) & $<$0.001\\ 
  2009 SE ($M_{1}=3$, $M_{2}=3$) & " & 0.042\\ 
    2009 SE  ($M_{1}=1$, $M_{2}=3$)& " & 0.41\\ 
2009 SE ($M_{1}=15$, $M_{2}=3$) &  U(0,0.6) $\times$ N($\mathbf{\mu}_{23}$,$\mathbf{\Sigma}_{23}$) & $<$0.001\\ 
  2009 SE ($M_{1}=3$, $M_{2}=3$) &  " & 0.028\\ 
  2009 SE  ($M_{1}=1$, $M_{2}=3$) & " & 0.22  \\  \noalign{\smallskip} \hline
\end{tabular}
\tablefoot{All tests have been carried out in ($\bar{d}$, $e$, $\sin{I}$) space.  The parameters $\mathbf{\mu}$ are $\mathbf{\Sigma}$ are the mean and covariance of the sample of $M$ asteroids. A "High" rating in the last column indicates that $p(M,m)$$>$0.5.}
\end{table*}
\begin{table*}
\caption{Osculating orbital elements of candidate cluster Mars Trojan asteroids at epoch JD2460600.5.}             
\label{table:oscelems}      
\centering          
\begin{tabular}{l@{ }rrrrrr}     
\hline\hline     \noalign{\smallskip}   
Designation & $a$ (au) & $e$ & $I$ (${}^{\circ}$)& $\varpi$ (${}^{\circ}$)& $\Omega$ (${}^{\circ}$)& $M$ (${}^{\circ}$)\\ 
\hline \noalign{\smallskip}  
\multicolumn{7}{c}{2009 SE Cluster }\\ \hline \noalign{\smallskip}
2009 $\mbox{SE}$ & 1.524459831 & 0.06496350736 & 20.62521595 & 1.007622369 & 6.803030083 & 358.6760928\\  
2015 $\mbox{TL}_{144}$ & 1.523918200 & 0.07799062229  & 19.60629849 & 34.33331021 & 351.8093779 & 273.6095801\\
 2016 $\mbox{AA}_{165}$ & 1.522908794 & 0.08951585534    & 18.72023279 & 40.59091644& 318.5245579 & 309.4527087\\ 
\hline \noalign{\smallskip}  
 \multicolumn{7}{c}{2018 $\mbox{EC}_{4}$ Cluster}\\ \hline \noalign{\smallskip}
  2011 $\mbox{SP}_{189}$ & 1.523719787 & 0.04038905315   & 19.89842888& 123.4137114& 0.6445263902 & 228.4729279 \\ 
    2018 $\mbox{EC}_{4}$ & 1.523544266 & 0.06042920722    & 21.83648685 & 31.63211779 & 47.34986354  & 321.9333061 \\ 
  2018 $\mbox{FM}_{29}$ & 1.523783588& 0.04730964899    & 21.49977945& 114.7875757& 167.9134320 & 240.1018711 \\   
 \hline \noalign{\smallskip} 
\end{tabular}
\tablefoot{Retrieved from {\tt AstDys} on 20 September 2024. All values reported to ten significant figures.}
\end{table*}
Within the Eureka family, we identify a grouping composed of 2011 $\mbox{SP}_{189}$, 2018 $\mbox{EC}_{4}$ and 2018 $\mbox{FM}_{29}$ (Fig.~\ref{fig:lr_vs_e_vs_i}) with $\delta\left[L\mbox{, }e\mbox{, }I\right]$=$0.38^{\circ}\mbox{$\pm$}0.21,0.0017\mbox{$\pm$}0.0021,0.18^{\circ}\mbox{$\pm$}0.14$ (errors reported at 2-$\sigma$ significance). Hereafter we refer to this grouping as the 2018 $\mbox{EC}_{4}$ cluster after its brightest member. We apply the MES compactness criterion (Section~\ref{sec:stats}) with a gaussian null distribution defined from the sample of $M=12$ Eureka family asteroids. The cumulative distribution function of MES volumes for 1,000 Monte Carlo trials and for all possible $m=3$ subsets is shown in the top panel of Fig.~\ref{fig:clustertests} where we find a $\sim$4\% probability that a grouping more compact than these three asteroids could have occurred by chance within the Eureka family. If we include nearby asteroid (311999) to this group and repeat the procedure, we find that this larger grouping is no longer significant ($p$-value $>$0.5). We therefore conclude that the $\mbox{EC}_{4}$ cluster is the largest statistically significant sub-grouping of family asteroids under the current observational completeness of the family. We return to the relation between this cluster and (311999) in Section~\ref{sec:origin}.

Turning now to Trojans outside of the Eureka family, we observe that the libration widths of two of the new stable Trojans highlighted in Table~\ref{table:propelems} are much higher than any Eureka family member but similar to that of 2009 SE, with the proper orbit of 2016 $\mbox{AA}_{165}$ in particular being almost identical to that of SE ($\delta\left[L\mbox{, }e\mbox{, }I\right]$=$0.36^{\circ}\mbox{$\pm$}0.96,0.0002\mbox{$\pm$}0.0021,0.23^{\circ}\mbox{$\pm$}0.14$). The third asteroid, 2015 $\mbox{TL}_{144}$, has the highest libration width of any stable Mars Trojan to-date, including (121514) at $\mbox{L}_{4}$ (Table~\ref{table:propelems}), but otherwise its eccentricity and inclination are similar to those of the other two asteroids ($\delta\left[e\mbox{, }I\right]$=$0.0027\mbox{$\pm$}0.0021,0.34^{\circ}\mbox{$\pm$}0.14$). 

The strongly elongated shape of this orbital grouping ($\sigma_{\bar{d}}$ $\gg$ $ \sigma_{e}$, $\sigma_{\sin I}$) renders the MES approach unsuitable and we resort to the CI-based approach to assess its statistical significance as a cluster of related asteroids. Application of the {\tt kmeans} algorithm with $k=2$ separates the population of $M_{1}+M_{2}=15$ $\mbox{L}_{5}$ asteroids into the candidate cluster and the Eureka family with a CI value $>$99.9\% significant against the null hypothesis (Table~\ref{table:sign}). To assess the robustness of the result, we replace the gaussian $\bar{d}$ variate with a uniform variate drawn between $0$ and $0.6$ and repeat the procedure. We then substitute the full Eureka family with only the 3 brightest family asteroids to force size parity between the two clusters, finding again that {\tt kmeans} separates the clusters with $>$95\% significance (Fig.~\ref{fig:clustertests}, middle panel). We conclude that identification of this grouping as a separate cluster is robust and therefore these small asteroids must be related in some way. For the remainder of the paper, we refer to the cluster by the name of its brightest member, 2009 SE.

The very close pair 2016 $\mbox{AA}_{165}$ - 2009 SE admits to investigation by the MES approach. Here we find that, as for the full cluster, statistical significance depends on the initial sample size $M$. For $M$$=$$15$ we obtain a significance of 91\% percent, however this assumption ignores the strong statistical association between Eureka family asteroids. Reducing the sample size to $M$$=$$6$ and then $M$$=$$4$ as done previously increases significance to 96\% and 99.5\% respectively, on a par with our earlier conclusions for either cluster. The case $M$$=$$6$ is shown in Fig.~\ref{fig:clustertests}, bottom panel.

As a final note in this Section, we point out that the asteroid sizes in these clusters are comparable to - though at the lower end of - members of identified pairs and clusters in the Main Belt \citep{VokrouhlickyNesvorny2008,Pravec.et.al2010,Pravec.et.al2018}.
However, unlike the Main Belt clusters that can be recognised by virtue of remarkably similar osculating orbits, Trojan cluster asteroids do not show this degree of similarity (Table~\ref{table:oscelems}) and are likely older than their Main Belt counterparts (see next Section). We postulate here that they have remained compact and identifiable by virtue of similar proper orbits owing, on one hand, to the overall sparsity of MT asteroids and, on the other, their spatial confinement $60^{\circ}$ ahead or behind Mars which mitigates differential orbital evolution by inverse-distance perturbations from the other planets \citep{Christou2013}.
\section{\label{sec:origin}Discussion: Cluster origin scenarios}
\subsection {2018 $\mbox{EC}_{4}$ cluster: a product of cascade disruption?}
Asteroids in both new groups occupy a relatively narrow range in $H$ implying similar-sized objects. Asteroid size can be estimated through the expression
\begin{equation}
D=\frac{1329}{\sqrt{p_{V}}} 10^{-H/5}
\label{eq:size_albedo}
\end{equation}
if the geometric albedo $p_{\rm V}$ is known.
Here we assume $p_{V}=0.2$ similar to other Eureka family asteroids \citep{Christou2013}.

For the $\mbox{EC}_{4}$ cluster we obtain nominal diameter estimates of 276m (2018 $\mbox{EC}_{4}$), 191m (2011 $\mbox{SP}_{189}$) and 175m (2018 $\mbox{FM}_{29}$), yielding same-density mass ratios in the range $0.25$-$0.77$, generally high for products of YORP-induced rotational fission \citep[mass ratio $\sim$0.2;][]{Pravec.et.al2010,Pravec.et.al2018}. 

Alternatively, the parent body of this cluster may be another Eureka family asteroid. The nearest family member is (311999) (Gig.~\ref{fig:lr_vs_e_vs_i}) with $D$$\sim$0.7 km. A genetic relationship with this asteroid would be more compatible with the rotational fission mechanism since, even if the group separated from (311999) as one single object, the inferred mass ratio of $0.094$ fits comfortably within the range predicted by theory. A relative youth for this cluster as implied by its apparent compactness is also consistent with separation from (311999). As this asteroid is itself thought to be a Eureka YORPlet, its link to the cluster implies an ongoing process of cascade disruption of the Eureka progenitor, similar to that invoked for some main-belt clusters \citep{Fatka.et.al2020}. 

By comparing the orbital distribution of numerically propagated test particles under the Yarkovsky effect to the observed distribution of Trojans, \citet{Cuk.et.al2015} obtained an approximate upper limit on the family age of 1 Gyr, corresponding to the separation event responsible for 2011 $\mbox{SC}_{191}$ (point at extreme bottom right in Fig.~\ref{fig:eureka_shape}).
We make use of that result here to estimate the age of the $\mbox{EC}_{4}$ cluster. We assume that the separation distance $s$ in $\bar{d}$ -- $\sin{I}$ space\footnote{i.e. ${ds}^{2} = d \left(\bar{d}\right)^{2} + d\left(\sin{I}\right)^{2}$} of the asteroid from the parent body after time $t_{\rm age}$ is related to the separation rate $\dot{s}$ by
\begin{equation}
s/t_{age}=\left(\dot{s}  - \dot{s}_{PB} \right)\mbox{.}
\label{eq:S1}
\end{equation}
 If the same inverse-linear size dependence holds for $\dot{s}$ as for the semimajor axis drift $\dot{a}$ in the absence of resonance, Eq.~\ref{eq:S1} becomes
\begin{equation}
s/t_{age}=\dot{s}\left(1  - D/D_{PB} \right)\mbox{.}
\label{eq:S2}
\end{equation}
From the analysis in Section~\ref{sec:eureka} we have $s_{SC191}$$\simeq$ 0.0536. Adopting $p_{V}=0.4$ for Eureka family asteroids as in \citeauthor{Cuk.et.al2015}, Eqs.~\ref{eq:size_albedo} and \ref{eq:S2} yield
\begin{equation}
\dot{s}_{SC191}=0.070\mbox{ Gyr}^{-1}
\end{equation}
and, by applying Eq.~\ref{eq:S2} to (311999) and each of the $\mbox{EC}_{4}$ cluster asteroids we obtain $t_{age}$= 44-127 Myr. As the Yarkovsky drift rate depends on a number of unknown bulk and surface characteristics of the asteroid \citep{Bottke.et.al2006}, the size-corrected value obtained here may differ from that of 2011 $\mbox{SC}_{191}$ while an additional uncertainty factor of $\sim$2 arises from the statistical uncertainty of the computed proper elements (Table~\ref{table:propelems}). However, even if the true separation ages are several times shorter than our estimated lower bound, this cluster appears to be significantly older than other previously identified clusters or pairs of small main-belt asteroids \cite[$\lesssim$4 Myr; ][]{Pravec.et.al2018,Pravec.et.al2019}.
\subsection {\label{sec:2009se}2009 SE cluster as the outcome of collisional fragmentation}
\begin{figure}
   \centering
   \includegraphics[width=\columnwidth]{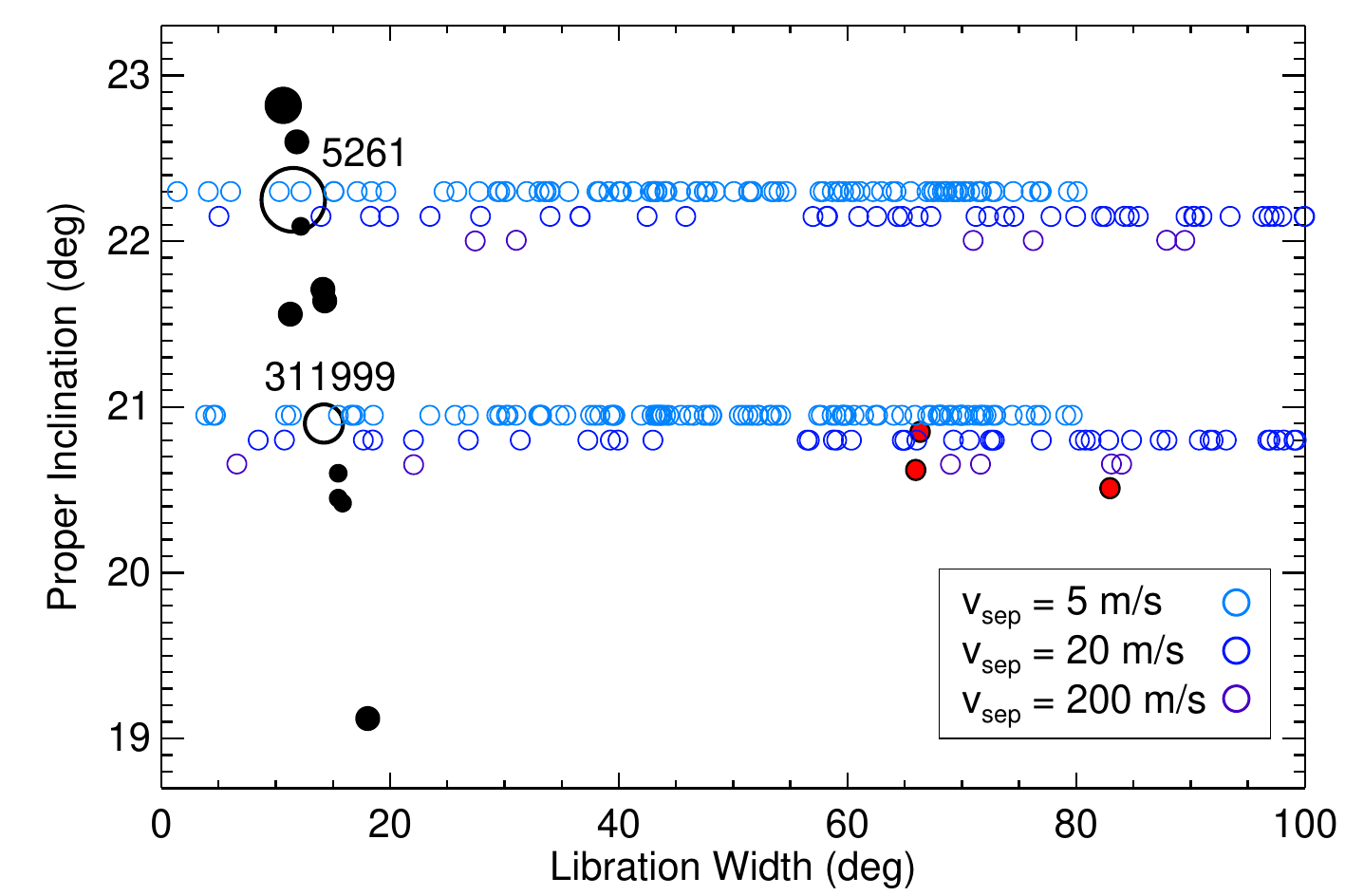}
   \includegraphics[width=\columnwidth]{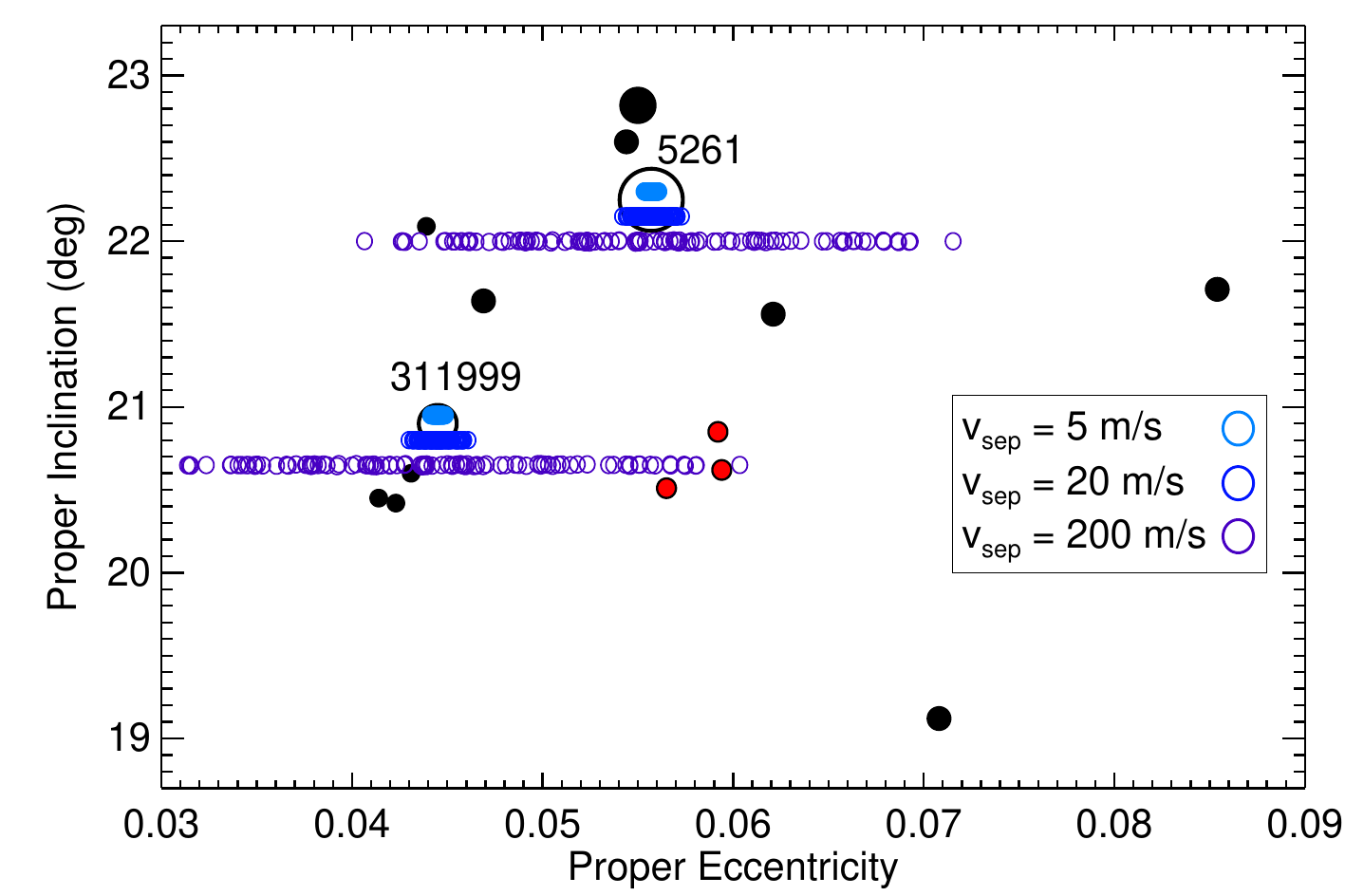}
      \caption{Orbital distribution of 100 particles (blue open symbols) that separated from asteroids (5261) and (311999) (black open symbols) with different speeds. Particle sets are staggered by $0.1^{\circ}$ in inclination for clarity. Red points correspond to members of the 2009 SE cluster.
      } 
         \label{fig:ejecta}
   \end{figure}
The similar eccentricity and inclination between this cluster and the Eureka family on one hand but much higher libration width on the other, could arise if this cluster is a remnant of an ejecta cloud from a past collision on a Mars Trojan. Such ejecta would escape from the surface of a parent body of diameter $D$ and bulk density $\rho$ if ${\rm v}_{\rm ej} > {\rm v}_{\rm esc}$ where \begin{equation}
{\rm v}_{\rm esc} = D \sqrt{(2/3)\mbox{ } G \pi \rho}  
\end{equation} and $G$ is the universal gravitational constant. The separation velocity of magnitude $ {\rm v}_{\rm sep}$=$\sqrt{{\rm v}^{2}_{\rm ej} - {\rm v}^{2}_{\rm esc}}$ for escaping fragments would cause differences between fragment and parent body orbits given by the Gauss equations \cite[e.g.][]{Nesvorny.et.al2006}:
\begin{eqnarray}
\label{eq:gauss_da}
\delta a / a \mbox{}&=& \frac{2 }{{\rm v}_{\rm circ} \left(1-e^2\right){}^{1/2}} \left[ \left( 1 + e\cos{f} \right) {\rm v}_{t} +  (e \sin{f}) \mbox{ } {\rm v}_{r} \right] \\
\label{eq:gauss_de}
 \delta e &=& \frac{\sqrt{1-e^{2}}}{{\rm v}_{\rm circ}} \left[\frac{e + 2 \cos{f} + e \cos^2{f}}{ 1 + e\cos{f}} {\rm v}_{t} + (\sin{f}) \mbox{ } {\rm v}_{r} \right]  \\
 \label{eq:gauss_di}
 \delta I &= &\frac{\sqrt{1-e^{2}}}{{\rm v}_{\rm circ}}\frac{\cos{\left(\omega+f\right)}}{1 + e \cos{f}}{\rm v}_{\rm w}
\end{eqnarray}
where the orbit elements are that of the parent asteroid, ${\rm v}_{\rm circ}=\sqrt{G M_{\odot}/a}$ is the heliocentric speed for a circular orbit of radius $a$ while ${\rm v}_{\rm t}$, ${\rm v}_{\rm r}$ and ${\rm v}_{\rm w}$ are the respective transverse, radial and out-of-plane components of $\mathbf{v}_{\rm sep}$. For $e$$\sim$$0$ and $a=a_{P}$, Eqs.~(\ref{eq:libration_width}) and (\ref{eq:gauss_da}) give 
\begin{equation}
 {\rm v}_{\rm sep, 0}\simeq 11.2 \mbox{ m}\mbox{ s}^{-1}
 \end{equation}
 for the minimum velocity ``kick'' needed to evict a small-amplitude ($\bar{d}$$\sim$$0$) Mars Trojan. A kick of this magnitude would, at the same time, have a negligible effect on the other elements ($\delta e$, $\delta I = \mathcal{O}(10^{-3}$) from Eqs.~\ref{eq:gauss_de} and \ref{eq:gauss_di}).
 Figure~\ref{fig:ejecta} shows the distribution of randomly-generated particles that separated from Eureka and (311999) with fixed separation speeds of 5, 20 and 200 m $\mbox{s}^{-1}$. Their orbits were obtained from Eqs~\ref{eq:gauss_da}-\ref{eq:gauss_di} by sampling $\omega$ and $f$ uniformly over the range $0$-2$\pi$ and for an isotropic separation direction. Values for the remaining elements were set to be those of the respective parent asteroids while $L$ was calculated from $\delta a$ using the method of \citet{Vokrouhlicky.et.al2024}.

We observe that a separation speed of 5 m $\mbox{s}^{-1}$ from either (5261) or (311999) is sufficient to increase $L$ to the values of the 2009 SE cluster (top panel). This scenario would be compatible with separation from (311999) in particular, except that an additional mechanism is then required to increase $e$ from 0.045 to $\sim$0.06 to match that of (311999) (bottom panel). Delivering such a change in fragment eccentricity from an impact on the asteroid requires a separation speed of 200 m $\mbox{s}^{-1}$, much higher than typical fragment ejection speeds from collisional disruption of km-sized bodies \cite[$\lesssim$10 m $\mbox{s}^{-1}$;][]{Jutzi.et.al2009}. 

An additional factor to consider is the timing of the impact event and subsequent orbital evolution of the fragments. We have seen \cite[Section~\ref{sec:eureka} and][]{Cuk.et.al2015} that small MTs would gradually migrate to orbits with lower inclination and higher libration width due to the Yarkovsky effect. Low-speed ejection of the 2009 SE cluster asteroids from (5261) sometime in the past would readily deliver the asteroids to their observed orbits without the need to invoke unusually high ejection speeds. Since the cluster's proper inclination is significantly lower from Eureka, the time elapsed since the collision event would be comparable to the Eureka family age. Assuming a similar rate of orbital migration and working as in Section 4.1, we find $t_{age}$$\simeq$$400$ Myr. 

The origin scenario we propose for this cluster is compatible with the one available  model of the likely collisional history of the Trojans \citep{Christou.et.al2017}, with about even odds that Eureka has experienced no catastrophic disruption \cite[$Q$$\sim$$Q^{\ast}$ where $Q$ is the impactor kinetic energy per unit target mass and $Q^{\ast}$ is an impactor- and target-dependent threshold; ][]{HousenHolsapple1990,BenzAsphaug1999}
over the age of the solar system. Impacts with $Q < Q^{\ast}$ should occur more frequently and we propose that at least one such event capable of imparting a speed of $\sim$5 m $\mbox{s}^{-1}$ to fragments $1/10$ the size of Eureka occurred within the most recent $\sim$1 Gyr of the solar system's history. 

Interestingly, this timeframe is also the expected collisional lifetime of D$\sim$300m MTs such as 2009 SE and 2018 $\mbox{EC}_{4}$ \citep{Christou.et.al2017}. The existence of either clusterings might therefore signify their relatively recent creation from progenitor bodies derived from Eureka, (311999) or both. 
In some cases the member separation epoch in young asteroid families may be constrained by backwards propagation of the orbits \citep{Nesvorny.et.al2002a,Carruba.et.al2018}. We relegate such an exercise for the MT families to future work.  
\section{\label{sec:conclusions}Conclusions}
\begin{itemize}
\item[--] We identify three new faint ($H$=$20$$-$$21$) $\mbox{L}_{5}$ Mars Trojans, raising the total number of trailing Trojan cloud residents to 16. Among the new finds is 2015 $\mbox{TL}_{144}$ with the highest width of tadpole-type libration ($\sim$$83^{\circ}$) among stable MTs.\\\vspace{-2mm}
\item[--] The new Trojans have similar eccentricity and inclination to the previously identified Eureka family of MT asteroids\citep{Christou2013,deLaFuenteMarcoses2013}. However, two of the objects have quite high libration width and may not be rotational fission products (``YORPlets'') as suggested for Eureka family asteroids \citep{Christou2013,Cuk.et.al2015,Christou.et.al2020}. We propose instead that these objects, together with 2009 SE identified previously by \citet{delafuentemarcos2021}, originated as collisional impact ejecta from a Eureka family asteroid with the most likely parent being (5261) Eureka itself. We also identify an additional statistically significant cluster of similarly faint Eureka family members that may have formed as YORPlets of asteroid (311999) sometime in the last $10^{8}$ yr.\\\vspace{-2mm}
\item[--] We revisit the significance of the correlation between libration amplitude and inclination attributed to Yarkovsky-driven orbital evolution \citep{Cuk.et.al2015} by calculating the Pearson correlation coefficient for the Eureka family asteroids. Our estimate of $94$\% implies a strong association between these two orbital properties. We further confirm the lack of Eureka-derived YORPlets with orbital evolution dominated by a positive Yarkovsky acceleration, consistent with the proposed dominance of the seasonal over the diurnal Yarkovsky effect over the $\sim$1 Gyr age of the family \citep{Cuk.et.al2015}. 
\end{itemize}
\vspace{-1mm}
While many families of $1$-$1,000$ km Main Belt asteroids have now been identified \citep{Nesvorny.et.al2015,Nesvorny.et.al2024}, the high MB asteroid number density and the overlap between families makes it difficult to identify small asteroid clusters older than $\sim$$10^{7}$ yr \citep{Pravec.et.al2018,Vokrouhlicky.et.al2024}. At the same time, MT proximity to Earth combined with the relative sparsity of their number and an environment free from frequent collisions and from planetary close approaches highlight this population as a unique natural laboratory to study small asteroid evolution.

The decadal {\it Legacy Survey of Space and Time} \citep{Jones.et.al2016} will discover and catalog approximately 5 million main belt asteroids in addition to the $\gtrsim$1 million currently known. Assuming similar discovery efficiency and a size distribution $n\left(<\right.$$\left.D\right)\propto D^{-0.45}$ for MTs \citep{Christou.et.al2020}, {\it LSST} should find $\sim$70 new MTs as faint as $H$$=$$23$ ($D$$\sim$$75$ m for a geometric albedo of 0.2), allowing to test the conclusions of this paper by comparing the orbital distributions of those smaller asteroids and of existing MTs. The outcome of this exercise will have wider implications for the efficiency of diurnal vs seasonal Yarkovsky and the relative roles of radiation forces vs collisions for sub-km asteroids on timescales $10^{7}$-$10^{9}$ yr. This is a regime where few observational constraints exist, yet extremely relevant to the important coupled problems of NEA/meteorite source regions and the migration routes of asteroid fragments to the Earth. 
\begin{acknowledgements}
ADO was supported by a grant of the Italian National Institute for Astrophysics for fundamental researches projects (INAF, act n.~38/2023). Work by A.~Humpage was supported by the Science and Technology Facilities Council (STFC). Astronomical research at the Armagh Observatory \& Planetarium is grant-aided by the Northern Ireland Department for Communities (DfC).
\end{acknowledgements}
%
\bibliographystyle{aa} 
\bibliography{new_mt_clusters} 
%


 
%
%

\begin{appendix} 
\section{\label{app:eureka}Statistical analysis of Eureka family proper orbits.}
\begin{table}
\caption{Proper element sample moments and covariance spectral decomposition for the 12 Eureka family asteroids.}    
\label{table:covelems}      
\centering          
\begin{tabular}{@{}rl@{}lll@{}}     
\hline\hline     \noalign{\smallskip}   

& Means &  \multicolumn{3}{c}{{\it Covariance} / {\bf correlation} matrices} \\
\hline \noalign{\smallskip}
 &  &  $\bar{d}$ &  $e_{P}$ & $\sin I_{P}$ \\
$\bar{d}$ & 0.016525 & \hspace{2mm}{\it 0.29033} & {\bf 0.0185723} & {\bf $-$0.939136}  \\
$e_{P}$ & 0.0537917 & \hspace{2mm}{\it 0.013475} & {\it 1.81315} & {\bf  \hspace{2mm}0.0248728} \\
$\sin I_{P}$ & 0.363952 & {\it $-$0.887128} & {\it 0.0587158} & \hspace{2mm}{\it 3.07344} \\ \hline     \noalign{\smallskip}  
  & Eigenvalues & \multicolumn{3}{c}{Eigenvectors }\\ \hline     \noalign{\smallskip}  
 & 0.0311209 & \hspace{2mm}0.959802 & $-$0.0164896 & \hspace{2mm}0.280193 \\
  & 1.81182 &  \hspace{2mm}0.0255079 &  \hspace{2mm}0.999266 & $-$0.0285696\\
  & 3.33398 & $-$0.279516 & \hspace{2mm}0.0345683 & \hspace{2mm}0.959518 \\ \noalign{\smallskip} \hline     
\end{tabular}
\tablefoot{All quantities reported to six significant digits. Covariance and correlation are displayed in lower diagonal and upper diagonal form respectively. Elements and eigenvalues of the covariance matrix have been multiplied by $10^{4}$.}
\end{table}
\section{\label{app:propunc} Uncertainty modelling of Mars Trojan proper element estimates}
\begin{table}
\caption{Proper element uncertainties of Mars Trojan asteroids.}             
\label{table:elemerrs}      
\centering          
\begin{tabular}{@{}rc@{ }c@{ }c@{ }c@{ }c@{ }c}     
\hline\hline     \noalign{\smallskip}   
 & $\sigma_{L\rm,o}$ & $\sigma_{L\rm,m}$ &  &  & $\sigma_{I\rm,o}$ & $\sigma_{I\rm,m}$ \\ 
Designation & (${}^{\circ}$) &  (${}^{\circ}$) & $\sigma_{e\rm,o}$ & $\sigma_{e\rm,m}$ & (${}^{\circ}$)&  (${}^{\circ}$)\\ 
\hline \noalign{\smallskip}               
(5261) & 2.5e-4 & 0.13 & 6.8e-8  & 1.5e-3 & 5.4e-6 & 1e-1\\ 
(101429) & " & 0.70 & "  & "  & " & " \\ 
(121514) & " & 0.60 & "  & "  & " & " \\
(311999) & " & 0.13 & "  & "  & " & " \\  
(385250) & "  & 0.13 & "  & "  & " & " \\  
2009 $\mbox{SE}$ & 3.5e-1 & 0.63 & 1.9e-5 & "   & 3.8e-4 & " \\
2011 $\mbox{SC}_{191}$ & 2.5e-4 &  0.17 & 6.8e-8 & "   & 5.4e-6 & " \\ 
2011 $\mbox{SL}_{25}$ & 9.8e-3 & 0.18 & 2.9e-6 & "   & 3.4e-4 & " \\
2011 $\mbox{SP}_{189}$ & 3.5e-1 & 0.14 & 1.9e-5 & "   & 3.8e-4 & " \\
2011 $\mbox{UB}_{256}$ & 2.5e-4 & 0.13 & 6.8e-8  & "   & 5.4e-6 & " \\
2011 $\mbox{UN}_{63}$ & " & 0.13 & " & "   & " & " \\
2015 $\mbox{TL}_{144}$ & 9.8e-3 & 0.76 & 2.9e-6 &"   & 3.4e-4 & " \\ 
2016 $\mbox{AA}_{165}$ & " & 0.63 & " & "   & " & " \\ 
2016 $\mbox{CP}_{31}$ &  " & 0.13 & " & "   & " & " \\ 
2018 $\mbox{EC}_{4}$  & 2.5e-4 & 0.15 & 6.8e-8 & "   & 5.4e-6 & " \\
2018 $\mbox{EC}_{4}$ &  9.8e-3 & 0.13 & 2.9e-6 & "   & 3.4e-4 & " \\ 
2018 $\mbox{FC}_{4}$  & 3.5e-1 & 0.12 & 1.9e-5 & "   & 3.8e-4 & " \\
2018 $\mbox{FM}_{29}$  & 9.8e-3 & 0.15 & 2.9e-6 & "   & 3.4e-4 & " \\ 
 \noalign{\smallskip}
\hline                  
\end{tabular}
\end{table}
   \begin{figure}
   \centering
   \includegraphics[width=\columnwidth]{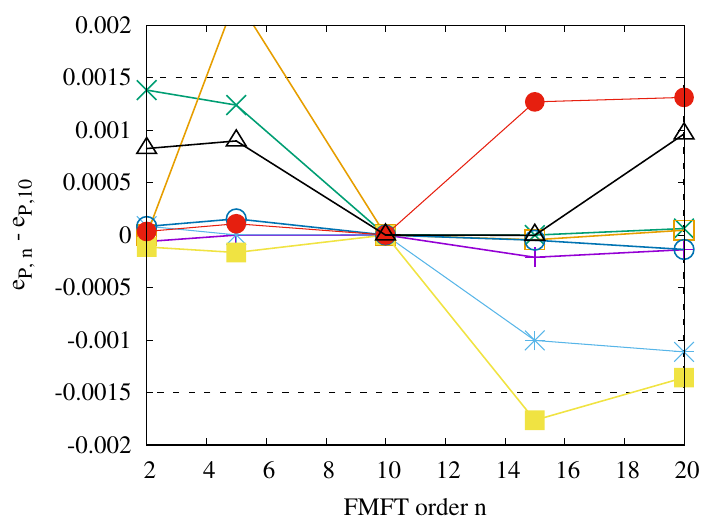}
   \includegraphics[width=\columnwidth]{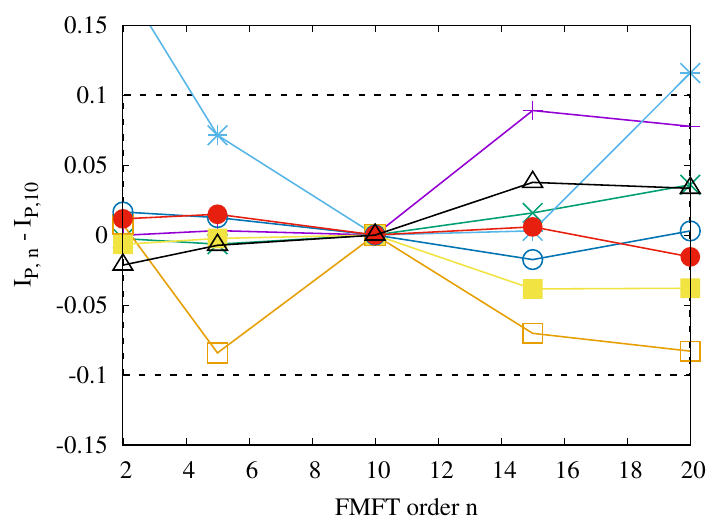}
      \caption{Estimates of proper eccentricity (top) and inclination (bottom) of MTs obtained by applying FMFT for different Fourier polynomial order. Displayed values are referenced to the case $n$$=$$10$. Dashed horizontal lines show the adopted formal uncertainties in this work.}
         \label{fig:eiunc}
   \end{figure}
Formal uncertainties reported in Table~\ref{table:propelems} are calculated as 
\begin{equation}
\sigma_{x}=\sqrt{\sigma^{2}_{x,\rm o}+\sigma^{2}_{x,\rm m}}    
\end{equation}
where $x = L\mbox{, }e_{P} \mbox{ or } {I}_{P}$, the first term under the square root refers to the orbit uncertainty and the second to the fitting model. Estimates of $\sigma_{x,\rm o}$ and $\sigma_{x,\rm m}$ for the MTs are shown in Table~\ref{table:elemerrs}.  

For the libration width we adopt as model uncertainty $\sigma_{L,m}$ the 2-$\sigma$ dispersion of the 120-sample estimates obtained from the numerical time series (Section~\ref{sec:proper}). To calculate the error introduced by the filtering procedure in estimating $\sigma_{e,m}$ and $\sigma_{I,m}$, we apply FMFT to selected MTs and for different Fourier polynomial order $n$ (2, 5, 10, 15 and 20). Fig.~\ref{fig:eiunc} shows these estimates shifted vertically to match the value at $n$$=$$10$. Consequently, we adopt the values $0.0015$ and $0.1$ containing 95\% of $e_{P}$ and $I_{P}$ estimates as respective model errors for all MTs.

To estimate the uncertainty arising from the finite knowledge of the orbit, the asteroids were first ranked according to the semimajor axis uncertainty $\sigma_{a}$, as the coorbital state is most sensitive to this element. This information was obtained from the state covariance for epoch JD2460600.5 from {\tt AstDys}. These values ranged between $\sim$$ 2$$\times$$10^{-9}$ au (5261) to $\sim $$2$$\times$$10^{-6}$ au (2009 SE). Each asteroid was then assigned to one of three bins demarcated by the values $\sigma_{a}=5\times 10^{-7}$ and $\sigma_{a}=5\times 10^{-8}$. We selected one representative for each bin and calculated proper elements for 20 dynamical clones of each and adopted 95\% confidence intervals from the 20-sample statistics as our estimates of $\sigma_{x,\rm o}$ for all asteroids in the respective bins. The bin representative asteroids in order of increasing $\sigma_{a}$ were (5261), 2009 SE and 2015 $\mbox{TL}_{144}$.
\end{appendix}
\end{document}